\documentclass[twocolumn,pre,showpacs,preprintnumbers,groupedaddress,amsmath,amssymb,floatfix]{revtex4}
\usepackage{graphicx}
\usepackage{amsmath}
\usepackage{amssymb}
\usepackage{color}

\begin{document}

\title{Quasiuniversal connectedness percolation of polydisperse rod systems}

\author{Biagio Nigro}\affiliation{LPM, Ecole Polytechnique F\'ed\'erale de
Lausanne, Station 17, CH-1015 Lausanne, Switzerland}
\author{Claudio Grimaldi}\affiliation{LPM, Ecole Polytechnique F\'ed\'erale
de Lausanne, Station 17, CH-1015 Lausanne, Switzerland}
\author{Peter Ryser}\affiliation{LPM, Ecole Polytechnique F\'ed\'erale de
Lausanne, Station 17, CH-1015 Lausanne, Switzerland}
\author{Avik P. Chatterjee}\affiliation{Department of Chemistry, SUNY College of Environmental
Science and Forestry, One Forestry Drive, Syracuse, N.Y. 13210}
\author{Paul van der Schoot}\affiliation{Theory of Polymers and Soft Matter,
Eindhoven University of Technology, P.O. Box 513, 5600 MB Eindhoven, The Netherlands}
\affiliation{Institute for Theoretical Physics, Utrecht University, Leuvenlaan 4, 3584 CE Utrecht, The Netherlands}

\begin{abstract}
The connectedness percolation threshold ($\eta_c$) and critical coordination number ($Z_c$) of systems of
penetrable spherocylinders characterized by a length polydispersity are studied by way of Monte Carlo simulations
for several aspect ratio distributions.
We find that (i) $\eta_c$ is a nearly universal function of the weight-averaged aspect ratio, with an approximate
inverse dependence that extends to aspect ratios that are well below the slender rod limit and (ii) 
that percolation of impenetrable spherocylinders displays a similar quasiuniversal behavior.
For systems with a sufficiently high degree of polydispersity, we find that $Z_c$ can become smaller than unity, in analogy
with observations reported for generalized and complex networks.
\end{abstract}


\maketitle

Idealized elongated objects such as perfectly rigid cylinders, spherocylinders and prolate spheroids are prototypical
models for a wide array of technologically relevant systems that include liquid crystals, nanocomposites based on
filamentous fillers as well as fiber-reinforced materials. Percolation phenomena involving dramatic increases in,
e.g., structural rigidity and electrical and thermal conductivities of composites
with increasing filler loading are currently of particular interest \cite{Kyrylyuk2008}.
These increases are caused by the formation of an infinite cluster of in some sense connected particles at the critical 
loading, i.e., the percolation threshold.

It has been established by analytical \cite{Balberg1984,Bug1985,Kyrylyuk2008} and numerical
\cite{Balberg1984b,Garboczi1995,Neda1999,Saar2002,Foygel2005,Schilling2007,Berhan2007,Ambrosetti2010a}
studies that for dispersions of sufficiently elongated objects of identical size and shape, i.e, ``monodisperse" objects,
the geometric percolation threshold expressed in terms of the critical volume fraction of particles is inversely
proportional to the aspect ratio of the filler particles. This property is exploited in the fabrication of conducting
polymeric composites with very low conducting filler contents. Depending on the production processes of the composites,
however, the filler particles almost invariably exhibit a pronounced polydispersity in both size and shape \cite{Wang2006,Beck2005}.
Although it represents
a possible factor behind huge quantitative discrepancies between theory and experiments \cite{Deng2009,Bauhofer2009},
such polydispersity has received relatively little attention in terms of
theoretical modeling until fairly recently \cite{Kyrylyuk2008,Otten2009,Chatterjee2010}, 
Achievement of a theoretical understanding of how the continuum percolation
of fibrous fillers is affected by polydispersity is thus key to the controlled design of a large class of 
composite materials for practical particle size and shape distributions.

Recent analytical results obtained from integral equation methods \cite{Otten2009} and from an heuristic mapping onto a 
generalized Bethe lattice \cite{Chatterjee2010} predict that in the slender rod limit, where the particles have asymptotically 
large values of the aspect ratio, the volume fraction at the percolation 
threshold is inversely proportional to the weight average $L_w=\langle L^2\rangle/\langle L\rangle$ of the rod lengths, where 
the brackets imply number averages over the distribution of rod lengths $L$.
This Letter presents Monte Carlo (MC) results for the percolation threshold of isotropically oriented
spherocylindrical particles with length polydispersity and having aspect ratios ranging from $\sim 1$ to 
several hundreds. We show that the percolation threshold of polydisperse, interpenetrable
spherocylinders is a nearly universal function of $L_w$ over the entire range of aspect ratios considered. In 
addition, the percolation threshold closely follows the predicted $1/L_w$
behavior even for particles with aspect ratios that are considerably smaller than the slender rod limit,
thus generalizing the current theory.

For systems of \emph{impenetrable} spherocylinders with fixed 
$\sqrt{\langle L^2\rangle}/D$, where $D$ is the diameter of the hard core,
we show that the percolation threshold is nearly independent of the length distribution.
Finally, we find that the critical coordination number per particle at the percolation threshold (denoted $Z_c$)
can be \textit{smaller} than unity for polydisperse systems. Although similar observations have been reported for a number
of complex networks and in systems of hyperspheres in high-dimensional spaces \cite{Wagner2006}, this finding is novel in the context of the continuum percolation of three-dimensional objects.

We generate isotropically oriented distributions of penetrable rods by randomly placing $N$ penetrable shero-cylinders
with a distribution of lengths $L$ and identical diameter $\delta$ within a cubic box with periodic boundary conditions and
side length $\mathcal{L}$.
As a measure of the concentration of the spherocylinders we shall use the dimensionless
density $\eta=\rho\langle v\rangle$, where $\langle v\rangle=(\pi/6)\delta^3+(\pi/4)\delta^2\langle L\rangle$ is
the number-averaged volume of the spherocylinders, $\langle L\rangle =\int \!dL L f(L)$ is the mean (number-averaged) rod length
for a given distribution $f(L)$ of lengths, and $\rho=N/\mathcal{L}^3$ is the number density of the particles \cite{note1}.

We consider two spherocylinders as being connected if they overlap geometrically. The percolation
threshold is identified by ascertaining the minimal diameter $\delta_c$ (for fixed value of $\rho$)  for which a cluster of
connected particles spans the entire cubic box. This definition is equivalent to the usual procedure of finding
a critical density $\rho_c$  of spherocylinders  with fixed diameter, and has the additional advantages of: (i) being
computationally more convenient, and (ii) allowing a more direct relation to the conductivity $\sigma$ of rods through
the critical distance approximation $\sigma\propto\exp(-2\delta_c/\xi)$, where $\xi$ is the tunneling decay
length \cite{Ambrosetti2010a}.

In the following we shall use the critical distance $\delta_{c0}$, defined as $\delta_{c0}=2/\pi\rho \langle L\rangle^2$,
as our unit of length for polydisperse systems of penetrable spherocylinders.
This quantity corresponds to the critical distance
obtained from the second virial approximation formula $\eta_c=(1/2)\delta_{c0}/L$ for the critical concentration of a system of
monodisperse spherocylinders with identical lengths $L\gg \delta_{c0}$ chosen to coincide with $\langle L\rangle$.

\begin{figure}[t!]
\begin{center}
\includegraphics[scale=0.57,clip=true]{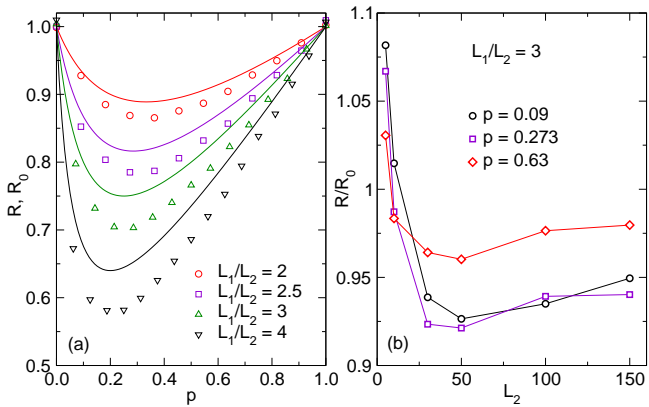}
\caption{(Color online) (a) critical distance ratio $R$ between polydisperse and monodisperse
spherocylinders as a function of the fractional occupancy $p$ of the longer rods
with shorter rod length fixed at $L_2=20$ in units of $\delta_{c0}=2/\pi\rho\langle L^2\rangle$ (see text). 
The solid lines represent Eq.~\eqref{ratio}. (b) The critical distance ratio $R$ in units of $R_0$ 
calculated from Eq.~\eqref{ratio} for $L_1/L_2=3$ as a function of the length $L_2$ of the shorter rods 
and for selected values of $p$.}\label{fig1}
\end{center}
\end{figure}

To find $\delta_c$ we employ the clustering method described in Ref.~\onlinecite{Nigro2011}, which allows
computation of the spanning probability as a function of the spherocylinder diameter $\delta$ for
fixed density $\rho$ (Supplemental Material \cite{SupMat}).
Figure~\ref{fig1}(a) shows the results obtained for polydisperse systems with a bimodal
length distribution $f(L)=p\delta(L-L_1)+(1-p)\delta(L-L_2)$ with $L_2=20$ and $L_1>L_2$, where $0\leq p \leq 1$ is the number
fraction of long rods. In the figure we display the ratio $R$ (symbols) of the critical distances
for the polydisperse rod system to those for monodisperse systems of spherocylinders with lengths
equal to $\langle L\rangle=\int dL Lf(L)$, which for the particular distribution
considered corresponds to $\langle L\rangle=pL_1+(1-p)L_2$.
The ratio $R$ of the critical distances is systematically reduced by
polydispersity and displays a minimum that becomes
deeper and moves towards smaller values of $p$ as $L_1/L_2$ is increased, implying that a small fraction of
longer rods can substantially lower the percolation threshold.

This trend is in full agreement with the theory of Ref.~\onlinecite{Otten2009}
based on the second virial approximation to the connectedness Ornstein-Zernike equation, which predicts for
$\langle L\rangle/\delta_c\gg 1$:
\begin{equation}
\label{etac}
\eta_c=\frac{1}{2}\frac{\delta_c}{L_w},
\end{equation}
where $L_w=\langle L^2\rangle/\langle L\rangle$ is the weight-averaged rod lengths. From
this equation, and by using $\eta_c\simeq \rho(\pi/4)\delta_c^2\langle L\rangle$ for $\langle L\rangle/\delta_c\gg 1$,
the critical distance is predicted to follow $\delta_c=2/\pi\rho\langle L^2\rangle$.
The reduction factor $R_0=\delta_c/\delta_{c0}$ predicted by the theory is thus
\begin{equation}
\label{ratio}
R_0=\frac{\langle L\rangle^2}{\langle L^2\rangle}=
\frac{(pL_1/L_2+1-p)^2}{p(L_1/L_2)^2+1-p},
\end{equation}
where the second equality applies for the bimodal length distribution. As shown in Fig.~\ref{fig1}(a)
the MC findings for $R$ are in semi-quantitative agreement with Eq.~\eqref{ratio} (solid lines),
although $R$ is consistently slightly smaller than $R_0$. This discrepancy could arise from the circumstance that
the values of $L_1$ and $L_2$ used in the simulations may be insufficiently large to achieve the slender
rod limit, which is a prerequisite for the validity of Eq.~\eqref{ratio}.
We examine this issue in Fig.~\ref{fig1}(b), which shows $R/R_0$ as a function
of $L_2$ (up to $L_2=150$
in units of $\delta_{c0}$) for $L_1/L_2= 3$ and selected values of $p$.
For $L_2\geq 20$, our MC results are less than $10$\% smaller than $R/R_0=1$.
Furthermore, for $L_2\geq 50$,
$R/R_0$ appears to increase monotonically (albeit somewhat slowly), which may indicate that the slender rod
length limit $R/R_0=1$ could ultimately be reached (for any $p$) only for very long rod lengths.

\begin{figure}[t]
\begin{center}
\includegraphics[scale=0.57,clip=true]{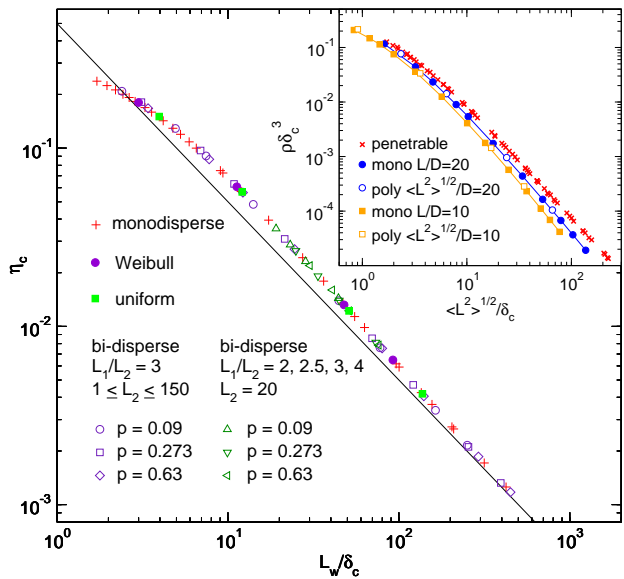}
\caption{(Color online) Critical reduced density $\eta_c$ as a function of $L_w/\delta_c$ for monodisperse (plus symbols),
bi-disperse (open symbols), Weibull (filled circles), and uniform (filled squares) distributions of spherocylinder 
lengths \cite{notedistr,SupMat}.
The solid line represents Eq.~\eqref{etac}.
Inset: $\rho\delta_c^3$ as a function of $\sqrt{\langle L^2\rangle}/\delta_c$ calculated for mododisperse and bidisperse
impenetrable spherocylinders with hard core diameter $D$. The \textsf{x} symbols are the results for both the monodisperse
and polydisperse penetrable rods of the main panel.}\label{fig2}
\end{center}
\end{figure}

The MC results shown in Fig.~\ref{fig1} and the relatively small deviations from Eq.~\eqref{ratio}
suggest that, for rods with identical radii, $L_w$ is the key quantity that controls
the percolation threshold for mixtures of penetrable spherocylinders.
This is demonstrated in Fig.~\ref{fig2} where $\eta_c$ is shown as a function
of $L_w/\delta_c$ for various bi-disperse (open symbols) and monodisperse ($+$ signs)
systems of spherocylinders (in the latter case $L_w$ is identical to the unique particle length).
Results for systems of spherocylinders for which the lengths follow  Weibull and 
uniform distributions are shown in Fig.~\ref{fig2} by filled circles and squares, respectively \cite{notedistr,SupMat}.
Surprisingly, all of our data collapse onto a single curve over the entire range of $L_w/\delta_c>1$,
implying that $\eta_c$ is a quasiuniversal function of $L_w/\delta_c$
independent of the particular distribution considered.

This finding is rather unexpected because the observed quasiuniversality extends well below the slender rod limit
of Eq.~\eqref{etac} (solid line), which is approached by the MC data
to within less than $10\%$ only for $L_w/\delta_c \gtrsim 200$.
Furthermore, even though Eq.~\eqref{etac} might be expected to apply only asymptotically for $L_w/\delta_c \gg 1$,
we observe that the data for  $L_w/\delta_c \gtrsim 10$ are well fitted by $a(L_w/\delta_c)^{-\beta}$
with $a=0.165\pm 0.009$ and $\beta\simeq 1.080\pm 0.002$.  The inverse scaling of $\eta_c$ with $L_w$ thus
applies approximatively even for spherocylinders with very modest aspect ratios.

We have also examined the effects of length polydispersity on the percolation of \emph{impenetrable} 
spherocylinders with identical hard-core diameters $D$.
Two impenetrable spherocylinders ($D\neq 0$) are considered to be connected if their surfaces approach closer 
than $\delta$. The percolation threshold for a given density $\rho$ of the particles is identified, as before, 
by the critical distance $\delta_c$.
In the slender rod limit the percolation threshold of impenetrable rods is predicted to follow $\phi_c=D^2/(2\delta_c L_w)$ \cite{Otten2009, Chatterjee2010},
where $\phi_c\simeq \rho(\pi/4)D^2\langle L\rangle$ is the critical volume fraction for the hard-core particles.
By noting that Eq.~\eqref{etac} is the percolation threshold for penetrable rods, and since
$\eta_c\simeq \rho(\pi/4)\delta_c^2\langle L\rangle$, we see that for sufficiently elongated rods the percolation relation
\begin{equation}
\label{uni2}
\rho\delta_c^3=(2/\pi)\delta_c^2/\langle L^2\rangle
\end{equation}
is predicted to be satisfied by both hard and penetrable rods, independent of their length distribution.

We have generated by MC simulations equilibrium dispersions of impenetrable spherocylinders with
different length ($L$) distributions.
The inset of Fig.~\ref{fig2} shows $\rho\delta_c^3$ as a function of $\sqrt{\langle L^2\rangle}/\delta_c$ for
monodisperse systems with $L/D=10$ and $20$ (filled symbols) and for two bidisperse cases with $L_1$, $L_2$, and $p$ chosen
as to give $\sqrt{\langle L^2\rangle}/D=10$ and $20$ (open symbols) \cite{SupMat}. 
Although for computational reasons the rod lengths considered by us are not large enough for
our results to fulfill Eq.~\eqref{uni2}, we see nevertheless that for a given $\sqrt{\langle L^2\rangle}/D$, $\rho\delta_c^3$ is
essentially independent of the particular rod length distribution. Furthermore, the calculated $\rho\delta_c^3$ values
for increasing $\sqrt{\langle L^2\rangle}/D$ tend to follow the same functional behavior of the interpenetrable spherocylinders
(\textsf{x} signs in the inset of Fig.~\ref{fig2}) \cite{noteuni}. This latter feature suggests that for sufficiently large $\sqrt{\langle L^2\rangle}/D$
there exists a universal relation of the form $\rho\delta_c^3=F(\sqrt{\langle L^2\rangle}/\delta_c)$,  which is expected to reduce to Eq.~\eqref{uni2}
for $\sqrt{\langle L^2\rangle}/\delta_c\gg 1$ and that applies to both penetrable and interpenetrable spherocylinders
over a wide range of $\sqrt{\langle L^2\rangle}/\delta_c$ values.

Although currently there is no theoretical explanation for the quasiuniversal dependence reported in Fig.~\ref{fig2},
a partial understanding may be achieved by following the
method developed in Ref.~\cite{Otten2009}. In this formalism, applied here for simplicity to
penetrable rods, the overall cluster size $S$ satisfies $S=\langle T(L)\rangle_L$
where $T(L)-\rho\langle\hat{C}^+(L,L',\delta_c)T(L')\rangle_{L'}=1$, and $\hat{C}^+(L,L',\delta_c)$ is the orientation-averaged  connectedness
direct correlation function at zero wave vector. Within the plausible ansatz $\hat{C}^+(L,L',\delta_c)=LL'c_{11}+(L+L' )\delta_c^2c_{10}+\delta_c^3c_{00}$
\cite{note2}, where the coefficients $\{c_{ij}\}$ are assumed to depend only upon the packing fraction,
it is found that for systems with different length distributions but equal values of $L_w/\delta_c$, $S$ diverges at
percolation thresholds that differ by $\sim \sigma_s^2\delta_c/L_w$ for $L_w/\delta_c\gg 1$ and by
$\sim (L_w/\delta_c)\sigma_s^2/(1+\sigma_s^2)$ for $L_w/\delta_c\ll 1$, where $\sigma_s^2=\langle L^2\rangle/\langle L\rangle ^2-1$
is the scaled variance.
Since the scaled variances for all length distributions considered in this work were always smaller than
$\sim 50\%$ \cite{SupMat}, for $L_w/\delta_c\gtrsim 10$
the expected deviation from universal behavior is thus only $\sim \sigma_s^2\delta_c/L_w
\lesssim 5\%$, which is consistent with the results of Fig.~\ref{fig2}.
Although we expect that systems with values of $\sigma_s^2$ much larger than those considered by us would imply a stronger
deviation from universality, $\sigma_s^2\lesssim 0.5$ is nevertheless representative of the scaled variances observed in
several real polydisperse systems of rod-like particles \cite{Wang2006,Beck2005}.

The quasiuniversal dependence of the percolation threshold upon $L_w$ implies a general \textit{non-universality} of
the critical coordination number $Z_c$, where $Z_c$ denotes the average number of contacts per
rod at the percolation threshold.
This is best viewed for the case of randomly placed and oriented overlapping objects for which
$Z_c=\eta_c\langle v_{\rm ex}\rangle/\langle v\rangle$, where
$\langle v_{\rm ex}\rangle$ is the excluded volume averaged over the orientations and the rod lengths. Given
that $\eta_c$ depends on $L_w/\delta_c$, while
\begin{equation}
\label{vex}
\frac{\langle v_{\rm ex}\rangle}{\langle v\rangle}=
8+3\frac{(\langle L\rangle/\delta_c)^2}{1+(3/2)\langle L\rangle/\delta_c}
\end{equation}
depends on the rod lengths through $\langle L\rangle/\delta_c$, we see that mixtures of rods with equal $\eta_c$
(\emph{i.e.}, equal $L_w$) may have rather different $Z_c$ if the distribution $f(L)$ of the rod lengths is such that
$\langle L\rangle\neq L_w$.

\begin{figure}[t]
\begin{center}
\includegraphics[scale=0.57,clip=true]{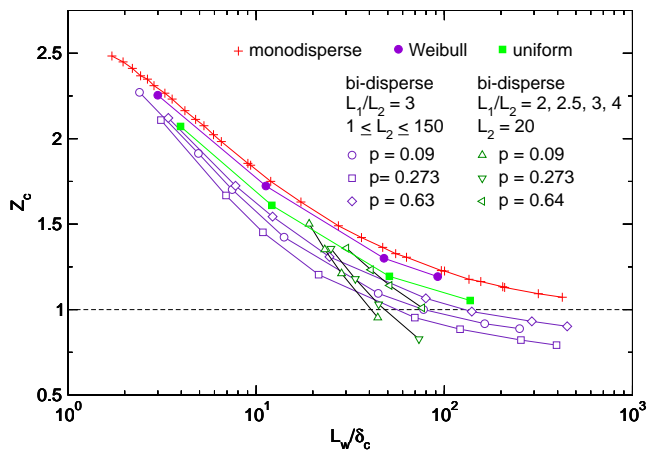}
\caption{(Color online) Critical coordination number $Z_c$ as a function of $L_w/\delta_c$ for polydisperse and
monodisperse spherocylinders. The symbols have the same meaning as in the main panel of Fig.~\ref{fig2}.}\label{fig3}
\end{center}
\end{figure}

Figure~\ref{fig3} shows the critical average number of connections per rod calculated from
$Z_c=\eta_c\langle v_{\rm ex}\rangle/\langle v\rangle$ for the same mixtures of penetrable spherocylinders considered in Fig.~\ref{fig2}.
We have verified that the configurational average of the connection per rods at $\delta_c$ coincides with
the excluded volume formula, as expected. We see that in general, $Z_c$ is sensitive to the extent of polydispersity,
although in the limit $L_w/\delta_c\rightarrow 0$ it can be expected that $Z_c$ should coincide with the result for
identical, overlapping spheres, namely $Z_c\simeq 2.74$ \cite{Baker2002}.
In particular, $Z_c$ is always larger than unity and approaches $Z_c\rightarrow 1$ asymptotically in the slender rod
limit for monodisperse systems.
In contrast, for distributions with sufficiently large values of the variance and of $L_w/\delta_c$,
polydisperse systems of rods may display fewer than one connection per particle at the threshold,
\emph{i.e.}, $Z_c <1$.

This latter feature is somewhat novel since the continuum percolation of objects
randomly dispersed in a three dimensional space is usually characterized by the condition $Z_c\geq 1$ \cite{Wagner2006}.
Indeed, to the best of our knowledge, percolation occurring with $Z_c<1$ has been reported only for penetrable
identical hyperspheres in spaces of dimensionality exceeding $12$ \cite{Wagner2006} and in random or complex networks
that are \textit{not} embedded in a physical space.
For example, given an uncorrelated network with nodes having a distribution of coordination numbers $z$, upon random
removal of nodes the network becomes disconnected at a critical node occupation probability
$p_c=\langle z\rangle/(\langle z^2\rangle-\langle z\rangle)$ \cite{Newman2003,Boccaletti2006}, which results from the
irrelevance of closed loops \cite{Cohen2000}.
The critical coordination number $Z_c=p_c\langle z\rangle$ is thus
\begin{equation}
\label{z}
Z_c=\frac{\langle z\rangle^2}{\langle z^2\rangle-\langle z\rangle},
\end{equation}
which can be smaller than unity when the node degree distribution is such that
$\langle z^2\rangle/\langle z\rangle-\langle z\rangle>1$.

The results of Fig.~\ref{fig3} that show that polydisperse rod mixtures may display $Z_c<1$ suggest that these
systems may relate to such classes of generalized graphs that can exhibit the same feature \cite{Silva2011}.
Indeed, as shown in Ref.~\onlinecite{Chatterjee2010}, Eq.~\eqref{z}
is also the critical coordination number of a generalized Bethe lattice that by construction lacks closed loops.
Hence, by applying the mapping $\langle z\rangle\rightarrow 2\langle L\rangle/\delta_c$,
$\langle z^2\rangle\rightarrow 4\langle L^2\rangle/\delta_c^2$ formulated for polydisperse slender rods in
Ref.~\onlinecite{Chatterjee2010}, we find
\begin{equation}
\label{zrods}
Z_c\rightarrow \frac{\langle L\rangle^2}{\langle L^2\rangle}=\frac{\langle L\rangle}{L_w}\leq 1,
\end{equation}
which is qualitatively consistent with the behavior of $Z_c$ seen in Fig.~\ref{fig3}.
A physical explanation for the observation that $Z_c < 1$ for sufficiently
polydisperse rod systems is provided by the fact that the percolating cluster
is predominantly comprised of the longer rods in the system, and the
shorter rods have a greater likelihood of being isolated \cite{Chatterjee2010}.
An interesting corollary arising from this interpretation is that, as in generalized
random networks where targeted removal of highly connected nodes enhances the percolation threshold \cite{Newman2003,Boccaletti2006},
preferential removal of the longer rods from the system may lead to similar enhancement of the critical concentration.

In conclusion, we have studied by MC simulations the effects of length polydispersity on the
percolation threshold of penetrable spherocylinders. 
We find a quasiuniversal dependence of the percolation threshold on $L_w$ that extends well
below the slender rod limit considered in Refs.~[\onlinecite{Otten2009,Chatterjee2010}].
For systems of impenetrable spherocylinders we find
that universality is fulfilled for a given $\sqrt{\langle L^2\rangle}/D$, where $D$ is the hard-core diameter.
The predicted quasiuniversality could be tested by experiments in systems 
of conducting fibrous fillers by altering the distribution of the rod lengths, e.g., by sonication, and measuring the
resulting change in the percolation threshold.
Furthermore we have demonstrated that the average number of connections per rod at the percolation threshold
can be smaller than unity for random distributions of rods that are sufficiently
slender and polydisperse.
This finding reveals an intriguing analogy with the case of random percolation in complex networks.

B. N. acknowledges support by the Swiss National Science Foundation (Grant No. 200020-135491).

\newpage

\section*{SUPPLEMENTAL MATERIAL}

\subsection{Calculation of the critical distance}
\label{critical}

\begin{figure}[b]
\begin{center}
\includegraphics[scale=0.55,clip=true]{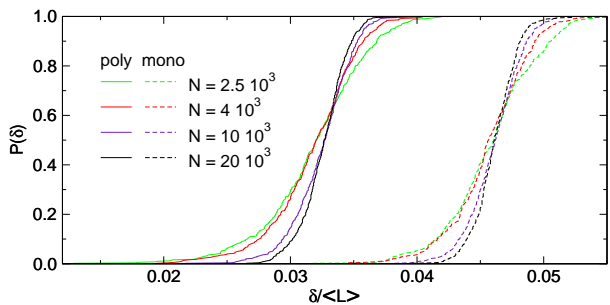}
\caption{Percolation probability $P(\delta)$ for different numbers $N$ of \emph{penetrable} spherocylinders
as a function of $\delta/\langle L\rangle$ for monodisperse (dashed lines) and bi-disperse (solid lines) spherocylinders.
The parameters of the bi-disperse distribution are $L_1=60$, $L_2=20$, and $p=0.21$. The monodisperse systems are generated
by considering rods with identical lengths coinciding with $\langle L\rangle$. All lengths are in units
of $\delta_{c0}=2/\pi\rho \langle L\rangle^2$. In the figure the number density is
fixed at $\rho=7.89\times10^{-4}$ for all cases.}\label{figS1}
\end{center}
\end{figure}

For both penetrable and impenetrable spherocylinders we follow same the method to calculate the critical distance
$\delta_c$. Namely, for a given number density $\rho$ of spherocylinders
which are either penetrable or impenetrable with hard-core diameter $D$, we coat each
spherocylinder with a penetrable shell of thickness $\delta/2$, and we consider two
spherocylinders to be connected if their penetrable shells overlap. For penetrable systems (i.e., for $D=0$)
$\delta$ represents the diameter of the penetrable spherocylinder.
For each realization of the system, we compute through the clustering method described in Ref.~\cite{Nigro2011supp}
the minimum value of $\delta$ such that a cluster of connected spherocylinders spans the entire sample.
By counting the number of instances that sample-spanning clusters appear for a given $\delta$, we construct the percolation
probability curve $P(\delta)$.

As in the main text, for systems of penetrable rods we adopt as unit of length the quantity
$\delta_{c0}=2/\pi\rho \langle L\rangle^2$, which corresponds to the critical distance in the second virial
approximation for monodisperse spherocylinders with length fixed at $\langle L\rangle$.
Examples of $P(\delta)$ obtained from $500$ realizations of polydisperse (solid lines)
and monodisperse (dashed lines) systems of penetrable spherocylinders are shown in
Fig.~\ref{figS1} for different numbers $N$ of spherocylinders with density fixed at $\rho=7.89\times10^{-4}$.
For the polydisperse cases we have considered a bi-disperse length distribution $f(L)=p\delta(L-L_1)+(1-p)\delta(L-L_2)$
with $L_1=60$, $L_2=20$, and the number fraction of long rods $p=0.21$. The monodisperse systems were generated
by spherocylinders of length equal to $\langle L\rangle=\int dL Lf(L)$, which for the particular distribution
considered corresponds to $\langle L\rangle=28.4$.
Figure \ref{figS1} reveals that the spanning probabilities for the bi-dispersed systems are shifted to lower
values of $\delta$ when compared to the $P(\delta)$ curves for the monodisperse case, indicating that the polydisperse
systems percolate at smaller volume fractions.

For both the polydisperse and the monodisperse cases the curves for the two highest values of $N$ intersect at
approximately $P=1/2$, which we take as our criterion for identifying the critical distance $\delta_c$. For the
particular case of Fig.~\ref{figS1} we find $\delta_c/\langle L\rangle\simeq 0.046$ and $0.032$ for the monodisperse
and polydisperse cases, respectively.

\begin{figure}[b]
\begin{center}
\includegraphics[scale=0.38,clip=true]{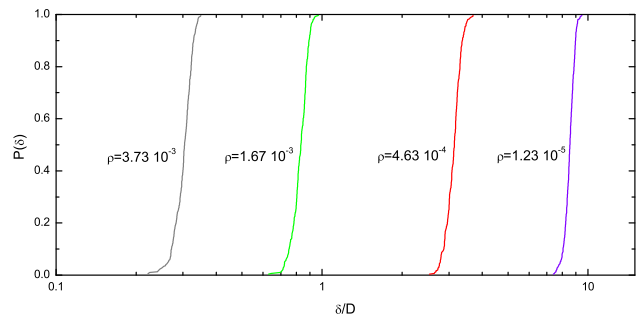}
\caption{Percolation probability $P(\delta)$ for different densities $\rho$ of \emph{impenetrable}
and bi-disperse spherocylinders as a function of $\delta/D$, where $D$ is the hard-core diameter. The parameters of the
bi-disperse distribution considered in the figure are $L_1/D=30$, $L_2/D=10$, and $p=3/8$. The corresponding value of
$\sqrt{\langle L^2\rangle}/D$ is $20$.}\label{figS2}
\end{center}
\end{figure}

Our results for penetrable spherocylinders shown in Figs. 1, 2, and 3 of the main text have been obtained by considering
simulation box sizes $\mathcal{L}$ such that $\mathcal{L}/L_1 \geq 5$, where $L_1$ is the largest rod length for any given
distribution, and the number $N$ of particles exceeds $2\times 10^4$. The resulting critical distances have been obtained
by adopting the criterion $P(\delta_c)=1/2$.

Figure \ref{figS2} shows the spanning probability $P(\delta)$ obtained from $300$ equilibrium configurations of bi-disperse systems
of impenetrable spherocylinders with $L_1/D=30$, $L_2/D=10$ and $p=3/8$ and for different values of the number density $\rho$.
From the largest to the lowest densities the number $N$ of spherocylinders decreases from $N=7000$ to $N=3000$ and the box
size $\mathcal{L}$ increases from $\mathcal{L}\simeq 4L_1$ to $\mathcal{L}\simeq 10 L_1$.
We have used a fitting to a simple sigmoidal function to evaluate the critical distance from $P(\delta_c)=1/2$ and from
the mean of the distribution function $dP(\delta)/d\delta$. The two methods give values of $\delta_c$ which differ at most by a
few percent.

\subsection{Bi-disperse, Weibull, and uniform distributions}
\label{distributions}

\begin{figure*}[t]
\begin{center}
\includegraphics[scale=0.5,clip=true]{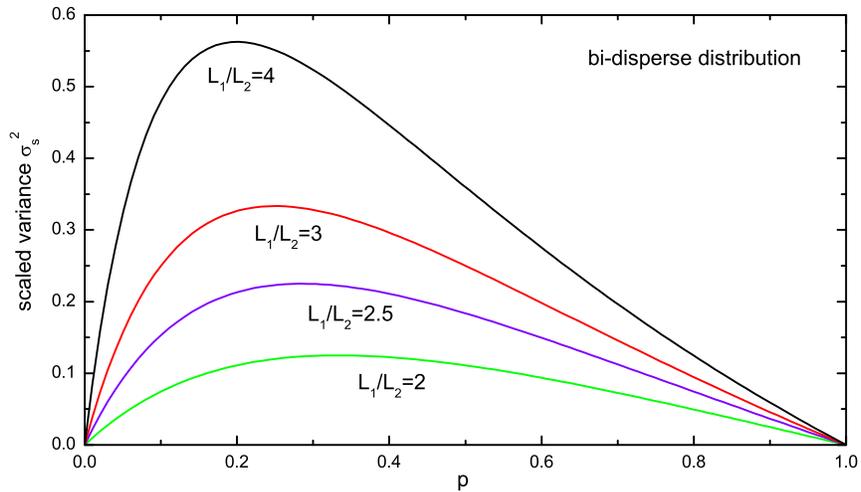}
\caption{Scaled variance $\sigma_s^2=\langle L^2\rangle/\langle L\rangle^2-1$ as a function of the number fraction $p$
of long rods for the bi-disperse distribution of lengths used in the calculations.}\label{figS3}
\end{center}
\end{figure*}

\begin{figure*}[t]
\begin{center}
\includegraphics[scale=0.5,clip=true]{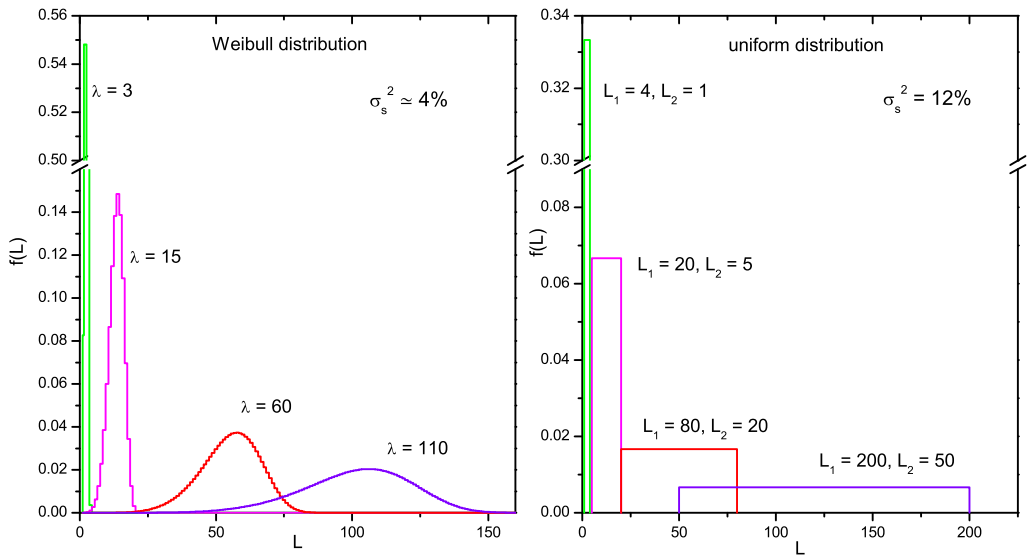}
\caption{Discretized Weibull (left panel) and uniform (right panel) distribution functions of the spherocylinder
lengths.}\label{figS4}
\end{center}
\end{figure*}

To avoid exceedingly large computational times, we have been careful to choose rod length distribution functions
with length $L_1$ of the longest rod not exceeding about $4$ times that ($L_2$) of the shortest one.
Despite of this constraint, we still have been able to generate length distributions with large scaled variances
$\sigma_s^2=\langle L^2\rangle/\langle L\rangle^2-1$. Among the different distributions considered, the bi-disperse
one, i.e., $f(L)=p\delta(L-L_1)+(1-p)\delta(L-L_2)$ with $0\leq p\leq 1$, had the largest $\sigma_s^2$ for a given
$L_1/L_2$ \cite{notesupp}. Figure~\ref{figS3} shows $\sigma_2^2=p(1-p)(n-1)^2/[p(n-1)+1]^2$ where $n=L_1/L_2=2$, $2.5$, $3$, and $4$
as a function of $p$. We see that for $p\sim 0.2$ and $L_1/L_2>3$, $\sigma_s^2$ is well above $30\%$.

In Fig.~\ref{figS4} we show the discretized Weibull (left panel) and uniform (right panel) distribution functions
of the rod lengths used in our study on the penetrable polydisperse spherocylinders.
The discretized Weibull distribution is defined as $f(L_i)=\exp[-(L_i/\lambda)^k]-\exp[-(L_{i+1}/\lambda)/^k]$,
where $L_i=i$ ($i=1,\, 2,\, 3, \ldots$) are the rod lengths \cite{weibull}. We have used $k=6$ and
$\lambda=3$, $15$, $60$, and $110$. The corresponding scaled variance
$\sigma_s^2=\langle L^2\rangle/\langle L\rangle^2-1$ is $\sigma_s^2\simeq 4\%$.
To guarantee that the ratio of lengths between the longest and shortest rods never exceeded $\sim 4$,
the distribution was truncated (and subsequently normalized) by eliminating from the sampling all
spherocylinders with $f(L_i)$ smaller than $10^{-2}$ of the maximum of the distribution.

Uniform distributions of rods (right panel of Fig.~\ref{figS4}) have been constructed from
$f(L)=1/(L_1-L_2)$ for $L_2\leq L\leq L_1$ and $f(L)=0$ otherwise, with $L_1/L_2=4$ and
$L_2=1$, $5$, $20$, and $50$. For all cases the scaled variance is $\sigma_s^2=12\%$.


\begin{thebibliography}{50}
\bibitem{note1}
The volume fraction $\phi$ for randomly placed penetrable spherocylinders is related to $\eta$ through
$\phi = 1-\exp(-\eta)$. Note that $\phi\simeq \eta$ for $\eta \ll 1$.
\bibitem{notedistr}
The systems with Weibull distribution have been generated by considering rods with lengths
$L_i=i$ ($i=1,\, 2,\, 3, \ldots$) distributed according to the discretized Weibull probability function $f(L_i)=\exp[-(L_i/\lambda)^k]-\exp[-(L_{i+1}/\lambda)/^k]$ with $k=6$ and
$\lambda=3$, $15$, $60$, and $110$. Uniform distributions of rods have been constructed from 
$f(L)=1/(L_1-L_2)$ for $L_2\leq L\leq L_1$ and $f(L)=0$ otherwise, with $L_1/L_2=4$ and 
$L_2=1$, $5$, $20$, and $50$.
\bibitem{noteuni}
When $\rho\delta_c^3$ is plotted as a function of $\sqrt{\langle L^2\rangle}/\delta_c$,
the system of penetrable rods displays a (quasi-)independence on the length distribution similar to that
observed from the $\eta_c$ versus $L_w/\delta_c$ plot of the main panel of Fig.~\ref{fig2}
\bibitem{note2}
The form of this ansatz is suggested by dimensional considerations and by the fact that the
direct correlation function is a local quantity.
The coefficients $c_{11}$, $c_{10}$, and $c_{00}$ represent respectively the contributions arising
from cylinder-cylinder, cylinder-cap, and cap-cap correlations. For $\eta\rightarrow 0$ they reduce
to the second virial approximation $c_{11}=\pi/2$, $c_{10}=\pi$, and $c_{00}=4\pi/3$.
\end{thebibliography}
\begin{thebibliography}{50}
\bibitem{notesupp}
For the penetrable case the rod lengths are in units of $\delta_{c0}$, while for impenetrable rods
the unit of length is the hard-core diameter $D$.
\end{thebibliography}
\end{document}